\documentclass[aps,prl,twocolumn,amsmath,amssymb]{revtex4}
\usepackage{bm}
\usepackage{graphicx}

\newcommand{\be}{\begin{eqnarray}}
\newcommand{\ee}{\end{eqnarray}}

\newcommand{\ba}{\begin{array}}
\newcommand{\ea}{\end{array}}

\begin{document}

\title{Security proof for cryptographic protocols based only on the monogamy of Bell's inequality violations}
\author{M. Paw{\l}owski}
\affiliation{Institute of Theoretical Physics and Astrophysics,
University of Gda{\'n}sk, PL-80-952 Gda{\'n}sk, Poland}
\begin{abstract}
We show that monogamy of Bell's inequality violations, which is strictly weaker condition than no-signaling is enough to prove security of quantum key distribution. We derive our results for a whole class of monogamy constraints and generalize our results to any theory that communicating parties may have access to. Some of these theories do not respect no-signaling principle yet still allow for secure communication. This proves that no-signaling is only a sufficient condition for the possibility of secure communication, but not the necessary one.  We also present some new qualitative results concerning the security of existing quantum key distribution protocols.
\end{abstract}
\maketitle

\openone
Quantum cryptography \cite{qkd-review}, or more precisely Quantum Key Distribution (QKD) is the first application of quantum information theory that has entered the commercial market. It's success stems from the fact that the safety of the key distribution protocol is based only on the laws of quantum mechanics, contrary to classical public key protocols which are secure only if certain task is complex enough for eavesdropper to perform. Though the laws of quantum mechanics have never been demonstrated to give incorrect predictions, physicists are constantly searching for more general theories, that could in principle counter these laws. In such theories quantum cryptography would no longer be secure unless the exact properties of quantum mechanics that warrant that security are also present in "post-quantum" world. One good candidate for such property is no-signaling principle. Since it appears not only in quantum mechanics but also in special relativity it is likely that in the future unification of these theories it will also hold. Indeed, it has been shown that no-signaling principle is sufficient  for the safety of quantum key distribution \cite{BHK}. That result significantly strengthens the QKD since, though it uses only the resources available in quantum theory it is safe in a whole family of theories that include but are not limited to quantum mechanics. In this paper we address the question, whether it is possible to find even weaker principle, which would lead to defining even larger group of theories for which quantum protocols remain secure. We show that the monogamy of Bell's inequality violations, which is strictly weaker condition than no-signaling is enough to prove the security of quantum key distribution. Therefore, for the first time, we are able to prove that quantum protocols are safe even if the eavesdropper has access to the resources of some {\it signaling} theories.

{\it Monogamy relations} - Let us define $\beta(\mathcal{X},\mathcal{Y})$ to be the left hand side of CHSH \cite{CHSH} inequality defined for parties $\mathcal{X}$ and $\mathcal{Y}$
\be \label{chsh}
\beta(\mathcal{X},\mathcal{Y})=\frac{1}{4}\sum_{x,y}P(X\oplus Y=xy|x,y)\leq \frac{3}{4}
\ee
where $X,Y\in\{0,1\}$ are the outcomes and  $x,y\in\{0,1\}$ are the settings of parties $\mathcal{X}$ and $\mathcal{Y}$ respectively. Throughout the whole paper we adopt notation in, which parties are denoted by "mathcal" letters, their settings by lowercase italic and outcomes by uppercase italic letters. One way to understand (\ref{chsh}) is to consider a game where two separated parties are being given by referees randomly chosen numbers - the settings $x,y$. Their goal is to return the outcomes $X,Y$ that with probability as high as possible satisfy $X\oplus Y=xy$. CHSH inequality states that there is no classical procedure that would allow them to have the success probability greater than $\frac{3}{4}$. On the other hand with quantum resources it is possible to have this probability as high as Tsirelson bound \cite{TSIRELSON}: $\frac{1}{2}\Big(1+\frac{1}{\sqrt{2}} \Big)$. This particular approach to CHSH inequality stresses the fact that the outcome of each of the parties can be the output of any procedure used by that party, as long as it does not violate any of the assumptions of the CHSH game. This will become very useful in the proof of our main result.

It has been shown \cite{ns-mon} (see also \cite{MAG}) that in any no-signaling theory, the following inequality must be satisfied
\be \label{monstd1}
\beta(\mathcal{A},\mathcal{B})+\beta(\mathcal{A},\mathcal{E})\leq\frac{3}{2}
\ee
We will refer to it as NS-monogamy. It is is weaker than no-signaling principle since NS-monogamy holds in any NS theory and the converse is not true. To see this let us imagine a theory which allows any three-partite of the form
\be \nonumber
P(A,B,E|a,b,e)\\ \label{prob}
=p P_{q_1}(A,B,E|a,b,e)+(1-p)P_{q_2}(A,E,B|a,e,b)
\ee
where $p\in [0,1]$ and
\be
P_{q}(X,Y,Z|x,y,z)=\frac{1}{2}P_q(X,Y|x,y)
\\
P_q(X,Y|x,y)=\left(\frac{1}{2}+(-1)^Yq\right)(X\oplus Y \oplus xy \oplus 1)
\ee
Probability distribution $P_q(X,Y|x,y)$ is a version of PR-box introduced in \cite{PR} with an additional parameter $q\in\left[0,\frac{1}{2}\right]$ which  describes the bias in the local probabilities. The original PR-box has $q=0$ and it is the only value of this parameter for which $P_q(X,Y|x,y)$ is no-signaling. So $P_{q}(X,Y,Z|x,y,z)$ is a probability distribution that allows for maximal algebraic violation of (\ref{chsh}) between $\mathcal{X}$ and $\mathcal{Y}$, while $\mathcal{Z}$ has uncorrelated white noise. The complete three-partite distribution $P(A,B,E|a,b,e)$ may be then understood as supplying, with probability $p$, parties $A$ and $B$ with a biased PR-box and with probability $1-p$ the parties $A$ and $E$. It is easy to check that this probability distribution is signaling for almost all combinations of $p,q_1$ and $q_2$ while satisfying (\ref{monstd1}).

Quantum mechanics satisfies not only (\ref{monstd1}) but also a stronger relation
\be \label{monstd2}
\Big(\beta(\mathcal{A},\mathcal{E})-\frac{1}{2}\Big)^2+\Big(\beta(\mathcal{A},\mathcal{B})-\frac{1}{2}\Big)^2\leq \frac{1}{8}
\ee
which has been proven in \cite{TV}. We will refer to it as QM-monogamy. The difference between (\ref{monstd2}) and the form given in \cite{TV} follows from the fact that here we use CHSH inequality written in the terms of probabilities, while in \cite{TV} it is written in the terms of correlation functions.

In order to be able to make more general statements we will express these (and any other) monogamy relations in a more homogenous form
\be \label{tmon}
\beta(\mathcal{A},\mathcal{E})\leq f^M_{T}(\beta(\mathcal{A},\mathcal{B}))
\ee
where $f^M_T$ is the function that describes the monogamy of the theory $T$. $f^M_{T}:\big[\frac{1}{2},1\big]\to [0,1]$ can be any non-increasing function. The domain of the function is chosen to be $\big[\frac{1}{2},1\big]$ since the form of CHSH inequality (\ref{chsh}) has algebraic bound equal to 1 and the value of $\frac{1}{2}$ corresponds to probability distribution describing white noise. We do not have to consider the behavior of $f^M_{T}$ below $\frac{1}{2}$ since every monogamy relation should also has the property that $f^M_{T}(\beta(\mathcal{A},\mathcal{B}))=f^M_T(1-\beta(\mathcal{A},\mathcal{B}))$ because if Bob always flips his outcome the value of $\beta(\mathcal{A},\mathcal{B})$ changes to $\beta'(\mathcal{A},\mathcal{B})=1-\beta(\mathcal{A},\mathcal{B})$. This cannot have any influence on $\beta(\mathcal{A},\mathcal{E})$. Therefore we can limit ourselves to the study of the monogamy for $\beta(\mathcal{A},\mathcal{B})\geq \frac{1}{2}$.
The above examples expressed in this way become
\be \label{nsmon}
\beta(\mathcal{A},\mathcal{E})\leq f^M_{NS}(\beta(\mathcal{A},\mathcal{B}))=\frac{3}{2}-\beta(\mathcal{A},\mathcal{B})
\\ \nonumber
\beta(\mathcal{A},\mathcal{E})\leq f^M_{QM}(\beta(\mathcal{A},\mathcal{B}))=
\sqrt{\frac{1}{8}-\Big(\beta(\mathcal{A},\mathcal{B})-\frac{1}{2}\Big)^2}+\frac{1}{2}
\\ \label{qmmon}
\ee

Now we are ready to present our main result, which states that (\ref{nsmon}) is sufficient condition for the security of quantum key distribution protocols against individual attacks.

{\it Main result} -  We will consider the following QKD protocol used by Alice and Bob: They share a large number of copies of a state, which they hope is close to singlet. For each copy they randomly choose one of the measurements that are optimal for the violation of CHSH inequality and write down the outcomes. Later they choose randomly a part of the runs and announce all the data corresponding to them to estimate $\beta(\mathcal{A},\mathcal{B})$. In the rest of the cases only Alice announces her choice of the basis. Her outcome is going to be the key. This is exactly the CHSH protocol, which security against individual attacks in no-signaling theories has been shown in \cite{ACM}. It has been proven \cite{CK} that Alice and Bob will be able to establish secure communication if
\be
I(\mathcal{A}:\mathcal{B})>I(\mathcal{A}:\mathcal{E}).
\ee
Since we are interested only in individual attacks and outcome of Alice is binary, the above condition simplifies to
\be \label{con}
P_B>P_E
\ee
where $P_B$ and $P_E$ are probabilities that Bob and Eve respectively guess the bit of Alice after the announcement of her choice of measurement basis. We will now show that if Eve is in the possession of a procedure that gives her high $P_E$, she could use the same procedure to play the CHSH game better than NS-monogamy allows.

Since the outcome of Alice $A$ equals $A=B\oplus ab$ with the probability $\beta(\mathcal{A},\mathcal{B})$, Bob, who knows $a,b$ and $B$ can guess that $A=B\oplus ab$ and be right with the probability
\be \label{pb}
P_B=\beta(\mathcal{A},\mathcal{B}).
\ee

Now, let us assume that Eve has a procedure that takes as an input the setting $a$ of Alice and generates $G$ as an output. The performance of this procedure is defined by the set of four probabilities $P_{ij}$ which describe the probability that $G$ equals the outcome of Alice if her setting was $j$ and Eve inputs $i$ to her procedure. For the reasons that will soon become clear we have to consider also the cases in which Eve inputs a value different than the setting of Alice, but when Eve is eavesdropping that will never happen. Her success probability is then
\be
P_E=\frac{P_{00}+P_{11}}{2}\leq\max\{P_{00},P_{11}\}.
\ee
We will consider two possible cases. First: $P_{00}\geq P_{11}$ and $P_E\leq P_{00}$. Let us see what value of $\beta(\mathcal{A},\mathcal{E})$ can Eve achieve with a strategy that involves the use of her procedure. Since in the CHSH game she does not know the setting of Alice, she will always input 0 to her procedure. If she returns as her outcome $E=G$ then the probabilities that appear in CHSH are equal
\be \nonumber
P(A\oplus E=0|a=0,e=0)=P_{00} \\ \nonumber
P(A\oplus E=0|a=1,e=0)=P_{01} \\ \nonumber
P(A\oplus E=0|a=0,e=1)=P_{00}
\\
P(A\oplus E=1|a=1,e=1)=1-P_{01}
\ee
That leads to
\be
\beta(\mathcal{A},\mathcal{E})=\frac{1}{2}P_{00}+\frac{1}{4}\geq\frac{1}{2}P_E+\frac{1}{4}
\ee

In the second case: $P_{00}< P_{11}$ and $P_E<P_{11}$. Now Eve adapts a different strategy. She will always input 1 and return $E=G\oplus e$ as her outcome. The probabilities that appear in CHSH are now
\be\nonumber
P(A\oplus E=0|a=0,e=0)=P_{10} \\ \nonumber
P(A\oplus E=0|a=1,e=0)=P_{11} \\ \nonumber
P(A\oplus E=0|a=0,e=1)=1-P_{10}
\\
P(A\oplus E=1|a=1,e=1)=P_{11}
\ee
That leads to
\be
\beta(\mathcal{A},\mathcal{E})=\frac{1}{2}P_{11}+\frac{1}{4}>\frac{1}{2}P_E+\frac{1}{4}
\ee
Therefore, for both cases, we have
\be \label{nec}
\frac{1}{2}P_E+\frac{1}{4}\leq \beta(\mathcal{A},\mathcal{E})
\ee
We now use NS-monogamy (\ref{nsmon}) and (\ref{pb}) to get
\be \label{pe}
\frac{1}{2}P_E+\frac{1}{4}\leq f^M_{NS}(P_B)
\ee
If we write $f^M_{NS}$ in the explicit form $f^M_{NS}(P_B)=\frac{3}{2}-P_B$, we get our main result that (\ref{pe}) implies (\ref{con}) as long as $P_B=\beta_{NS}(\mathcal{A},\mathcal{B})>\frac{5}{6}$. It is worth noticing that $\frac{5}{6}$ is less than the Tsirelson bound, so the correlations between $\mathcal{A}$ and $\mathcal{B}$ required for the security based only on monogamy are within reach of quantum mechanics. This result is similar to the conditions for security based on no-signaling derived in \cite {ACM} (see also \cite{Nicolas} for the conditions for CHSH protocol with pre-processing).

Note that (\ref{pe}) holds in every theory that has NS-monogamy. This theory can even be signaling like the example (\ref{prob}). This proves that no-signaling is only a sufficient condition for the existence of secure cryptography, not a necessary one.

{\it Generalization to other monogamies} - Note that the derivation of our main result does not depend on the explicit form of $f^M_{NS}$, which is only used later to establish the minimal value of $\beta(\mathcal{A},\mathcal{B})$ required for security. Since nowhere in the paper we assume no-signaling, we do not have to limit ourselves to the monogamy constraint derived from that principle.
If instead of NS-monogamy (\ref{nsmon}) we would assume QM-monogamy (\ref{qmmon}) (which can also hold in signaling theories), instead of (\ref{pe}) we would get
\be
\frac{1}{2}P_E+\frac{1}{4}\leq f^M_{QM}(P_B),
\ee
which leads to weaker requirement: $\beta_{QM}(\mathcal{A},\mathcal{B})>\frac{1}{2}+\frac{1}{\sqrt{10}}$.

In general, for any theory $T$ which has T-monogamy defined by (\ref{tmon}) the following condition holds
\be \label{pt}
\frac{1}{2}P_E+\frac{1}{4}\leq f^M_{T}(P_B)
\ee
It will be possible to construct a safe cryptographic protocol for Alice and Bob who have access only to quantum resources as long as the highest value of $P_B$ achievable by quantum mechanics causes (\ref{pt}) to bound $P_E$ to values lower than $P_B$. Rewriting (\ref{nec}) as
\be
P_E\leq 2\beta(\mathcal{A},\mathcal{E}) -\frac{1}{2}
\ee
and keeping in mind that $P_B=\beta(\mathcal{A},\mathcal{B})$ we find that the condition (\ref{con}) will be satisfied if the T-monogamy can imply
\be
\beta(\mathcal{A},\mathcal{B})>2\beta(\mathcal{A},\mathcal{E}) -\frac{1}{2}
\ee
or in other words
\be \label{suffi}
f^M_{T}\Big(\beta(\mathcal{A},\mathcal{B}) \Big)<\frac{1}{2}\beta(\mathcal{A},\mathcal{B})+\frac{1}{4}
\ee
Which, if Alice and Bob are able to reach the Tsirelson bound, becomes
\be \label{s2}
f^M_{T}\Bigg(\frac{1}{2}\Big(1+\frac{1}{\sqrt{2}}\Big)\Bigg)<\frac{1}{2}\Big(1+\frac{1}{2\sqrt{2}}\Big)
\ee
Inequality (\ref{suffi}) is the sufficient condition on T-monogamy to be able to warrant the safety of quantum key distribution protocol.

\begin{figure}[ht]
\includegraphics[scale=0.85]{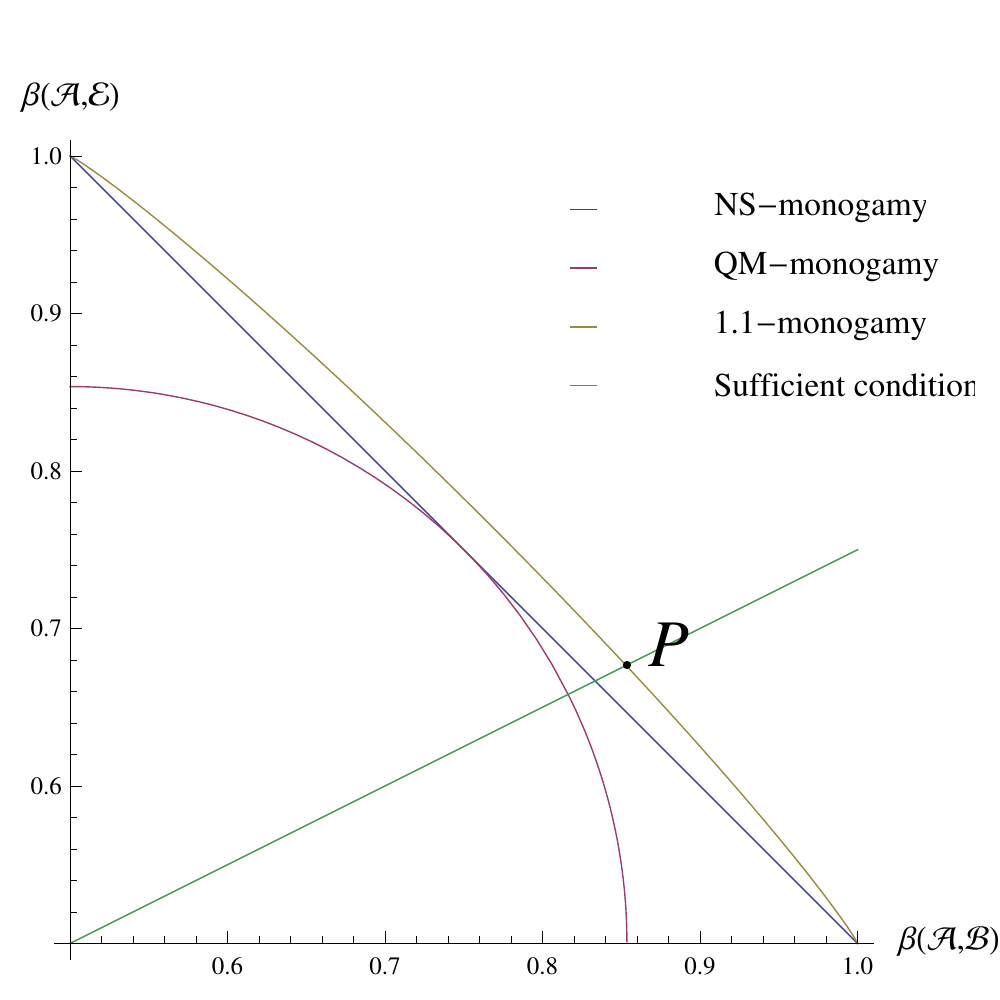}
\caption{Monogamy relations for different theories are plotted. The sufficient condition for the security of the key distribution protocol (\ref{suffi}) is described by the line that passes through $P=\Big(\frac{1}{2}\big(1+\frac{1}{\sqrt{2}}\big),\frac{1}{2}\big(1+\frac{1}{2\sqrt{2}}\big)\Big)$.
The intersection of this line with monogamy relation for any specific theory T gives the critical value of $\beta_T(\mathcal{A},\mathcal{B})$ above which Alice and Bob can have secure communication. More explicitly, if $\mathcal{A}$ and $\mathcal{B}$ estimate their $\beta(\mathcal{A},\mathcal{B})$ to be greater than $\beta_T(\mathcal{A},\mathcal{B})$, they can have secure communication against individual attacks of the eavesdropper who has access to the resources of theory T.
Point $P$ has its horizontal coordinate equal to the Tsirelson bound, so for any theory T that intersects the sufficient condition line before $P$, it is possible for $\mathcal{A}$ and $\mathcal{B}$ to have a quantum protocol secure against attacks from T regime.  Note that intersection of 1.1-monogamy with sufficient condition line is slightly before $P$. Therefore, quantum mechanics allows QKD secure against attacks by eavesdroppers from all three theories.
}
\label{mons}
\end{figure}

For example, let us consider a 1.1-monogamy which is weaker than that of any no-signaling theory. It is defined by
\be
f^M_{1.1}(\beta)=\frac{1}{2}\Big(\big(1-(2\beta-1)^{1.1}\big)^\frac{1}{1.1}+1\Big)
\ee
The critical value of $\beta(\mathcal{A},\mathcal{B})$ for theory with 1.1-monogamy is 0.8530 which, though quite high, is still 0.07\% lower than Tsirelson bound.

All three discussed monogamies along with the sufficient condition (\ref{suffi}) are presented in fig. \ref{mons}.

Our result can be easily generalized to the the case where Alice and Bob have access to the resources of theory T, while the eavesdropper has access to the resources of T'. If T allows for $\beta_{T}(\mathcal{A},\mathcal{B})$ and T' has T'-monogamy then (\ref{s2}) becomes
\be
f^M_{T'}\Big(\beta_{T}(\mathcal{A},\mathcal{B})\Big)<\frac{1}{2}\beta_{T}(\mathcal{A},\mathcal{B})+\frac{1}{4}
\ee

{\it Discussion} - In the proofs of security that are based either on laws of quantum mechanics or no-signaling principle there appears a critical value of $\beta(\mathcal{A},\mathcal{B})$. If Alice and Bob estimate their correlations to be below that value they cannot assume security. If they are above, then they are secure and the actual value of $\beta_{crit}(\mathcal{A},\mathcal{B})$ implies only the key rate. On the other hand, if we assume only monogamy than the higher $\beta(\mathcal{A},\mathcal{B})$ the larger the number of theories in which $\mathcal{A}$ and $\mathcal{B}$ are secure. That is an interesting corollary as it shows that cryptographic systems that use Bell inequalities to warrant the security not only get higher key rates as the violation of Bell's inequality raises but also gain more qualitative security as the number of theories against attacks from which they are protected raises too.

We have shown that it is possible to base the security of quantum key distribution only on monogamy relations for violations of Bell's inequality. That is significantly weaker assumption than the best previously known - no-signaling principle. Moreover, our proof does not depend on the specific type of the monogamy relation, but on its value at a single point:  $f^M_{T}\Big(\frac{1}{2}\big(1+\frac{1}{\sqrt{2}}\big)\Big)$. We have also generalized our results to give predictions for any theory that the communicating parties may have access to. We have proved that no-signaling is only a sufficient condition for the existence of secure cryptography, not a necessary one.

In the case of security based on no-signaling, the proofs went a long way from partial version presented in \cite{BHK} to universally-composable \cite{UC} shown in \cite{MAS}. We express hope that our paper will begin the similar path for security based on monogamy, which would strengthen quantum cryptography even further. We also conjecture that it is possible to prove universally-composable security based solely on monogamy.

{\it Acknowledgments} - I would like to thank Micha\l{} Horodecki and Nicolas Brunner for discussions. The work is  supported  by EC Project QAP and by polish research network LFPPI.

\end{document}